\documentclass{emulateapj}

\shorttitle{Mid-IR Resolution of 3 AU Disk Around Zeta Lep}
\shortauthors{Moerchen et al.}

\begin{document}

\title{Mid-Infrared Resolution of a 3 AU-Radius Debris Disk Around Zeta~Leporis\altaffilmark{1}}

\author{M. M. Moerchen, C. M. Telesco, C. Packham, and T. J. J. Kehoe}
\affil{University of Florida, 211 Bryant Space Science Center, P. O. Box 112055, Gainesville, FL 32611-2055; margaret@astro.ufl.edu}

\altaffiltext{1}{Observations were obtained at the Gemini Observatory, operated by AURA, Inc., under agreement with the NSF on behalf of the Gemini partnership: NSF (US), PPARC (UK), NRC (Canada), CONICYT (Chile), ARC (Australia), CNPq (Brazil), and CONICET (Argentina).}

\begin{abstract}
We present subarcsecond-resolution mid-infrared images of the debris disk surrounding the 230~Myr-old A star {\rm$\zeta$}~Lep.  Our data obtained with T-ReCS at Gemini South show the source to be unresolved at 10.4~$\mu$m but clearly extended at 18.3~${\rm\mu}$m.  Quadratic subtraction of the PSF profile from that of $\zeta$~Lep implies a characteristic radius for the dust disk of 3~AU, which is comparable in size to our solar system's asteroid belt.  Simple models suggest that the 18~$\mu$m flux is well approximated by two contiguous annuli of mid-infrared-emitting dust from 2--4 and 4--8~AU with a 3:1 flux ratio for the annuli, respectively. We consider two scenarios for the collisions that must be resupplying the dust population: (1) continuous ``steady state'' grinding of planetesimals, and (2) an isolated cataclysmic collision. We determine that radiation pressure and subsequent collisions are the dominant determinants of the disk morphology in either case, and that Poynting-Robertson drag is comparatively insignificant.
\end{abstract}


\section{Introduction}
Excess infrared flux associated with main sequence stars usually indicates the presence of thermally emitting dust in a circumstellar disk.  When the stellar age is long compared to the timescales for removal of pristine (i.e., primordial) material by Poynting-Robertson (P-R) drag or radiation pressure, the observed dust must be produced by collisional cascades of large particles and planetesimals or by cometary deposits \citep{bac92, wya99}.  Such disks are referred to as debris disks.  The A-type main-sequence star $\zeta$ Lep (HD~38678) was identified as a debris disk candidate by the IRAS discovery of its infrared excess \citep{aum91}.   It is one of several stars where the disk particles are significantly warmer than is typical for debris disks \citep{cot87,aum91}, implying that the particles are relatively close ($<$10~AU) to the star. For example, the $\sim$320~K mid-infrared color temperature of $\zeta$~Lep (Chen \& Jura 2001; this work) contrasts with archetypes such as the disk of $\beta$~Pic, which has a mid-infrared color temperature of $\sim$160~K for dust 80 AU from the star \citep{tel88}.  Furthermore, $\zeta$~Lep is one of the few such stars with warm disks that are close enough to us (21.5~pc, Perryman et al. 1997) that ground-based observers can reasonably expect to spatially resolve the disk.  Two recent observations of the disk drew our attention to the $\zeta$~Lep disk and motivated our program of mid-infrared imaging at Gemini: a strong constraint on the size of the disk at 18~$\mu$m at Keck (Chen \& Jura 2001, hereafter CJ01) and a {\em Spitzer} MIPS photometric survey of A-type stars showing that the infrared excess of $\zeta$~Lep is high among stars of comparable age \citep{rie05}.  CJ01 have noted that the disk radius must be smaller than 9 AU, and therefore probably comparable in size to our own asteroid belt, which makes it unique among debris disks discovered so far.  As we report here, we have imaged $\zeta$~Lep with Gemini to constrain more tightly the disk size and further examine dust production mechanisms.

\section{Observations}

	We obtained mid-infrared images of $\zeta$ Lep (HD~38678) on 2005 February 3 (UT) at Gemini South (Program ID GS-2005A-Q-2) with T-ReCS (Thermal Region Camera and Spectrograph), using the broadband N ($\lambda_{\rm c}$~=~10.36~$\mu$m, $\Delta\lambda$~=~5.27~$\mu$m) and narrowband Qa ($\lambda_{\rm c}$~=~18.30~$\mu$m, $\Delta\lambda$~=~1.51~$\mu$m) filters.   The total field of view is 29" x 22".  The data were reduced with the Gemini IRAF package.  The flux standards $\epsilon$~Lep and $\alpha$~CMi were observed before and after, respectively,  the group of target and PSF observations.  Flux measurements of $\zeta$~Lep at N (2.03~$\pm$~0.19~Jy) and Qa (0.96~$\pm$~0.06~Jy) are consistent with previous results \citep{faj98,che01}.  We observed a point-spread-function (PSF) comparison star, HD~42042, both before and after the target observations in each filter.  HD~42042, an M star located 6.6$^{\rm \circ}$ away from $\zeta$~Lep, is $\sim$10 times brighter than $\zeta$~Lep at 10.4~$\mu$m and $\sim$6 times brighter at 18.3~$\mu$m.  We divided the PSF and target integrations into pairs of nodsets to examine the potential effect of airmass on the full-width at half maximum (FWHM) value, and there is no direct correlation.

\section{Results}

	\subsection{Source Size \& Model}
	
	In the final stacked images, we find that the azimuthally averaged profiles of $\zeta$~Lep and the PSF star are identical at 10.4~$\mu$m, whereas  the source is clearly extended at 18.3~$\mu$m (Fig. 1).  We have partly characterized this extension by comparing the full width half-maximum (FWHM) of the profiles determined from a Moffat fit generated in IRAF.  At 10.4~$\mu$m, the fits to the PSF star and $\zeta$~Lep profiles measure 0.31"~$\pm$~0.02".  At 18.3~$\mu$m the profile fits give 0.54"~$\pm$~0.02" and 0.60"~$\pm$~0.02", respectively.  Even though we consider the FWHM radius of the profiles, we have determined those values from analytic fits to the full profiles. For $\zeta$~Lep, the indicated uncertainty is the standard deviation of the mean of the series of FWHM measurements made throughout the integration set, as described in greater detail below.    For the PSF star, we have only six FWHM measurements, and the uncertainty given is half of the excursion between the maximum and minimum values. Like CJ01, we do not resolve $\zeta$~Lep at 10.4~$\mu$m, but we do resolve the source at 18.3~$\mu$m (Fig.~1).

	To test the robustness of the latter result, we examined the FWHM of the PSF star and $\zeta$~Lep throughout the integration sequence.  Pupil rotation, incorrect guiding correction, and changes in the quality of seeing during long observations on a program object can result in a final image degraded by lower-frequency components  that are not accurately represented in the PSF determined from generally shorter integration times.  Thus, we have divided the total 18.3~$\mu$m integration sequence into bins of 120~s of on-source integration time and have measured the width of the Moffat profile fit for each of these subdivided images.  The displacement between the mean profile width of the PSF and the mean profile width of the source implies that the source is resolved.  Quadratic subtraction of the derived FWHM value of the PSF star from that of $\zeta$~Lep, as determined from the 18.3~$\mu$m images, indicates that the average disk radius is 0.14"~$\pm$~0.02", or 3.0~$\pm$~0.3 AU. CJ01 estimate a best-fit value of 2.15~AU, but they note that the uncertainties in their observations allow for values as large as 9~AU (CJ01; C. Chen, priv. comm.).  Our observations made under nearly diffraction-limited conditions imply that the size of the emitting region is indeed comparable to that of the asteroid belt in our solar system.
		
To constrain the geometry of the $\zeta$~Lep disk, we generated 2-D models of the expected 18.3 $\mu$m brightness distribution with a range of disk parameters.  Subtraction of the PSF image normalized to the peak brightness from the image of $\zeta$~Lep shows a residual ring of emission that, given the asymmetries evident in the PSF itself, is consistent with the disk being approximately azimuthally symmetric. We therefore assume a face-on disk geometry for our models and calculations, but this assumption has no effect on our conclusions.  (Our observations provide no constraint on the disk orientation.)  

Our models consist of a central source (the star) with either one or two annuli of dust emission each with uniform surface brightness, where the constant total flux determined by observations at 18.3~$\mu$m was equally divided between the central source and dust emission according to the estimated excess above the photosphere (CJ01).  For the single-annulus models, the free parameters of annulus width (2, 4, or 6~AU) and annulus inner edge distance from the star (2, 3, 4, 6, or 8~AU) are combined to yield fifteen models.  We also generated models with two concentric and contiguous annuli with different but uniform brightnesses.  In all of these models, the inner annulus extends from 2 to 4~AU, whereas the outer annulus extends from 4~AU to either 6~AU or 8~AU. The ratios of the total fluxes from the inner and outer annuli are either 1:1 or 3:1, respectively, resulting in four double-annuli models. The double-annulus is a simple geometric construct that likely approximates a more complex radially-dependent brightness.  The models were then convolved with a 2-D representation of the PSF that was generated from the azimuthally averaged PSF star observation at 18.3~$\mu$m.  A $\chi$-squared minimization comparison of the resulting model profiles to the source data shows that the source is best approximated by a model of two annuli.  In this model, a 2-4~AU annulus emits 75\% of the thermal dust flux, and a 4-8~AU annulus emits the remaining 25\%.  Our models are not exhaustive but do illustrate the nature of satisfactory fits to the observed profiles.  We also note that the dust emitting the unresolved 10.4~$\mu$m excess must lie interior to the resolved 18.3~$\mu$m flux and may be migrating inwards by P-R drag from the region of replenishment by collisions through a much lower-density region close to the star.
		
	\subsection{Particle Size, Mass and Lifetimes}

			The expected temperature for blackbody dust particles at $3.0\pm0.4~$AU from the star is 320$^{+24}_{-19}$~K.  This is comparable to the color temperature of 323$^{+27}_{-30}$~K inferred from the two observed mid-infrared flux densities and associated uncertainties.  Thus, the particles may be blackbody emitters, in marked contrast to mid-infrared emitting grains in other debris disks; e.g., dust 50~AU from $\beta$~Pic is at a temperature of $\gtrsim$140~K compared to the blackbody temperature of $\sim$65~K \citep{wei03,tel05}. Blackbody behavior requires particle radii of a few microns or larger.  Additional evidence that the particles orbiting $\zeta$~Lep are large is provided by recent mid-infrared spectroscopy of the source taken with the {\em Spitzer} Infrared Spectrograph (IRS) \citep{che06}.  Silicates are commonly found in circumstellar disks, but the spectrum of $\zeta$ Lep shows no evidence of a distinct silicate emission feature.   Silicate particles as large as 1-2~$\mu$m can result in a broad silicate feature, which is expected to disappear as particle radii increase above a few microns and the particles become optically thick in the mid-infrared (e.g., Przygodda et al. 2003).   \citet{che06} conclude that, if silicates are present, the majority of particle radii are larger than 10~$\mu$m.

			We estimate the size of the smallest particles expected to remain on bound orbits ($\beta$~=~$F_{\rm r}/F_{\rm g}$~$<$~0.5, Burns et al. 1979) in response to the forces of gravity and radiation pressure, assuming 14~L$_\odot$ and 1.9~M$_\odot$ for the stellar luminosity and mass, respectively, and unity for the radiation pressure efficiency.  Adopting the standard silicate density value of 2.5~g~cm$^{-3}$, we find that the minimum radius is 3.4~$\mu$m.  Particles smaller than this are blown out of the system by radiation pressure within a few orbital timescales.   The Keplerian orbital period at 3~AU from $\zeta$ Lep is $\sim$3~years, so grains smaller than the blowout radius of 3.4~$\mu$m should be removed on a decade timescale.  With this minimum particle size threshold, we estimate the number-density-weighted average dust particle radius of 4.7~$\mu$m by assuming that the incremental number of grains per radius interval $da$ is proportional to $a^{-3.5}$, which describes the equilibrium size distribution of the dust particle population in a collisional cascade \citep{doh69}.  
			
			Assuming that all particles have the average radius 4.7~$\mu$m and are in a thin shell 3.0~AU from the star, we estimate the dust mass from the amount of ultraviolet and optical flux reprocessed as thermal emission (Jura et al. 1995, Eq. 5).  With values for the fractional infrared luminosity $L_{{\rm IR}}$/L$_{\rm \ast}$~=~1.7~$\times~10^{-4}$ \citep{faj98} and particle density 2.5~g~cm$^{-3}$, we find a total dust mass of 6.7~$\times~10^{21}$~g, or 0.37\% of the total mass of the asteroids in our solar system. This value is $\sim$40\% of the mass estimate in CJ01 due to the smaller disk size (and therefore hotter dust) determined with our observations.  A change in the particle-size power-law exponent from 3.5 to 6.4 lowers the mass estimate by $\sim$20\%.

			Large dust particles ($a\gtrsim3.4~\mu$m) are subject to loss by spiraling into the star under P-R drag \citep{bur79}.  However, this process is only relevant if the time interval between destructive particle collisions is much longer than the P-R drag timescale \citep{wya05}.  The characteristic collisional lifetime of the average radius particle at a distance $r$ from the star is $t_{{\rm coll}}(r) = t_{{\rm per}}(r)/(4 \pi \tau_{{\rm eff}}(r))$, where $t_{{\rm per}}$ is the Keplerian orbital period and $\tau_{{\rm eff}}$ is the effective optical depth.  We estimate the effective optical depth, defined as the ratio of the total cross-sectional area of dust to the total area of the disk \citep{wya99}, from the total area of  average-sized (4.7~$\mu$m) particles necessary to produce the infrared luminosity and the area within the 2--8~AU annulus; for $\zeta$ Lep, this value is 2~$\times~10^{-4}$, approximately equal to the observed value for $L_{{\rm IR}}$/L$_{\rm \ast}$.
	Assuming a constant particle density, the ratio of the P-R drag timescale to the collisional timescale is a weak function of distance from the star, so we expect the relative significance of the dominant process to be similar throughout the disk.
	For example, the expected time for a particle to migrate by P-R drag from 6~AU to 3~AU (the center of each annulus) is $\sim$16000~y, several times longer than the collisional timescale at 4 AU (the nominal division between the annuli) of only $\sim$2000~years.  Therefore, we consider it unlikely that the particles orbiting $\zeta$~Lep migrate a substantial distance toward the star before being broken up in a collisional cascade.  We conclude that initially radiation pressure and, subsequently, collisions are the two dominant processes defining the population and distribution of particles in $\zeta$~Lep's disk.
 
\section{Dust Production Scenarios}

			A plot of infrared excess versus age \citep{rie05} of 266 A-type stars shows that $\zeta$~Lep has a high infrared excess compared to other main-sequence stars of similar age.  In this paper we have adopted the \citet{son01} value of 231~Myr, estimated with photometry-determined stellar parameters to fit to stellar evolutionary tracks; this estimate also incorporates the effects of rapid rotation, and a survey of A-type stars shows that $\zeta$~Lep's position in the H-R diagram is close to that of other stars with comparable age estimates \citep{jur98}.   
However, if the age is 100~Myr or less (cf. Barrado y Navascu\'{e}s 1998; Song et al.), the excess level for the disk is within the range of values typical for those age bins.  Therefore, the  dust production scenario that we consider most probable depends, to some extent, on the age we assume for $\zeta$ Lep: if the disk is older than 100~Myr (cf. Lachaume et al. 1999; Song et al. 2001), the excess is indeed anomalously high and may indicate a recent cataclysmic collision, but, if the disk is much younger, we may be observing the dust production and decay associated with a relatively steady collision rate of smaller bodies.  We examine both of these scenarios below.
			
Recent cataclysmic collisions may account for some of the excess emitting dust around $\beta$~Pic \citep{tel05} and Vega \citep{su05}, and may also contribute to the range of excesses observed for stars of similar ages \citep{rie05}.  The destruction of one planetesimal of radius $\sim$85 km could account for the entire observed mid-infrared-emitting dust mass in $\zeta$~Lep. To assess the plausibility of a single cataclysmic collision of such a body resulting in a disk resembling our best-fit model, we consider the dynamics of the dust particles immediately following their production in such an event. The initial orbital evolution of the newly formed dust particles is dominated by radiation pressure, which acts to  increase the semi-major axes of their orbits.  If we assume that these particles are produced by the destruction of a body on a circular orbit at 2~AU (the inner edge of the inner annulus in the best-fit model to our observations), we can estimate the radius for particles expected to migrate to a given distance, using standard relations (e.g., Kortenkamp \& Dermott 1998; Wyatt et al. 1999).  Since smaller particles will have larger apastron distances under the influence of radiation pressure, the size of a particle that can migrate to a given apastron distance corresponds to a lower limit on the size of particles interior to that location.  The radius of particles expected to reach 4~AU (5.1~$\mu$m) will be both the upper limit for particles in the 4--8~AU annulus and the lower limit for the 2--4~AU annulus, where in the latter case all smaller grains have been blown farther out.  The radius of particles expected to reach 8 AU (3.9~$\mu$m) sets the lower limit for particle radii in the outer 4--8~AU annulus.
			
		To characterize the particle distribution that would yield the flux ratio (3:1) between the annuli of the model fit to our observations ($\S$3.2), we sum over the areas of all the individual emitting particles, as described by a power-law particle distribution, assuming the radii are $>$5.1~$\mu$m in the inner annulus and 3.9--5.1~$\mu$m in the outer annulus.  (Since the radii of the particles that are expected to remain on bound orbits within 8 AU are larger than a few microns, we assume that the particles are blackbody emitters.) To estimate their temperature and flux, we then assume that all of the particles in the inner and outer annuli are at a characteristic distance of 3 and 6 AU from the star, respectively.   We find that, to match the model of our observations, the size distribution of the dust population resulting from a collision at 2~AU must be described by a power law with an exponent of 6.4, much steeper than the \citet{doh69} value of 3.5.  Indeed, there is no reason to expect the immediate post-collision population to resemble the Dohnanyi distribution that is thought to apply to a population that has evolved to collisional equilibrium \citep{dim90,dur99}. The post-collision distribution presumably depends heavily on the details of the collision and the composition of the parent body.  For example, with regard to the latter,  ``rubble piles" and differentiated asteroids may produce different relative amounts of small and large particles \citep{gro01}.  A particle size distribution consistent with our double-annulus model therefore suggests that the proportion of small particles produced in the hypothetical collision is larger than that implied by the Dohnanyi distribution.

		In contrast to the catastrophic collision scenario, dust production by relatively ``steady-state" grinding collisions may be more likely if the age of $\zeta$~Lep is less than 100~Myr and the excess level is not high compared to other stars of comparable age.  Indeed, CJ01 propose that dust is replenished in $\zeta$~Lep in an asteroid belt undergoing such a process.  They estimate the total mass of the asteroids constituting the parent-bodies' mass by assuming that the observed dust mass will, on average, be lost and replenished within the P-R lifetime of the particles.  If this has occurred continuously throughout the star's life, the required parent-body mass, $M_{\rm d}\times t_{{\rm age}}/t_{{\rm PR}} $,
			 is $\gtrsim$4~$\times~10^{26}$~g, which is $\sim$5\% of the primordial mass of asteroids ($\sim1~M_\earth$, e.g., Lissauer 1987) in our solar system.  However, since we contend that collisional cascades and subsequent radiation pressure effects are the dominant regulators of the dust population, we reconsider the estimate above for the more relevant collisional timescale $t_{{\rm coll}}$ rather than that of P-R drag and determine a mass of $\sim$6~$\times~10^{26}$~g, $\sim$10\% of the solar system's initial asteroidal mass.

\section{Discussion}
			
 There is a growing body of evidence for cataclysmic collisional events as a major, or even predominant, source of dust production in debris disks, as discussed in \S4.  A plot of simulated dust production rate in our solar system  \citep{der02} shows a slow decrease of dust area over time, punctuated by spikes from dust-producing collisions.   This illustrates that the amount of dust alone cannot indicate how long it has been since the most recent cataclysmic collision. Much greater spatial resolution that permits the particle orbits to be constrained is needed for us to better understand the collisional history of the $\zeta$~Lep debris disk.	The apparently restricted spatial extent of this disk may be due to planetary perturbations such as resonant trapping, a scenario that has also been proposed for the debris disk around the K star HD~69830. This disk's limited extent was inferred by {\em Spitzer} photometry \citep{bei05} and may be due to the influence of its triple planet system \citep{lov06}.  The presence of planets may also assist in redistribution of the dust azimuthally by secular perturbations, resonant interactions, gravitational scattering, or accretion.

		Our key conclusion is that we have resolved the disk of $\zeta$ Lep at 18.3~$\mu$m, which has a characteristic size comparable to that of our own asteroid belt. This is the only spatially resolved debris disk of this size, and it presents a new disk archetype that contrasts with Kuiper Belt-like disks extending tens to hundreds of AU from the central star.

\acknowledgments

We thank Christine Chen and Michael Jura for helpful comments. M. M. M. acknowledges that this work was performed [in part] under contract with 
JPL funded by NASA through the Michelson Fellowship Program. JPL is managed for NASA by the California Institute of Technology.  C. M. T. acknowledges support from NSF grant AST-0098392.

{\it Facilities:}  \facility{Gemini:South (T-ReCS)}.

\clearpage
			 
\begin{figure}
\includegraphics[width=\columnwidth]{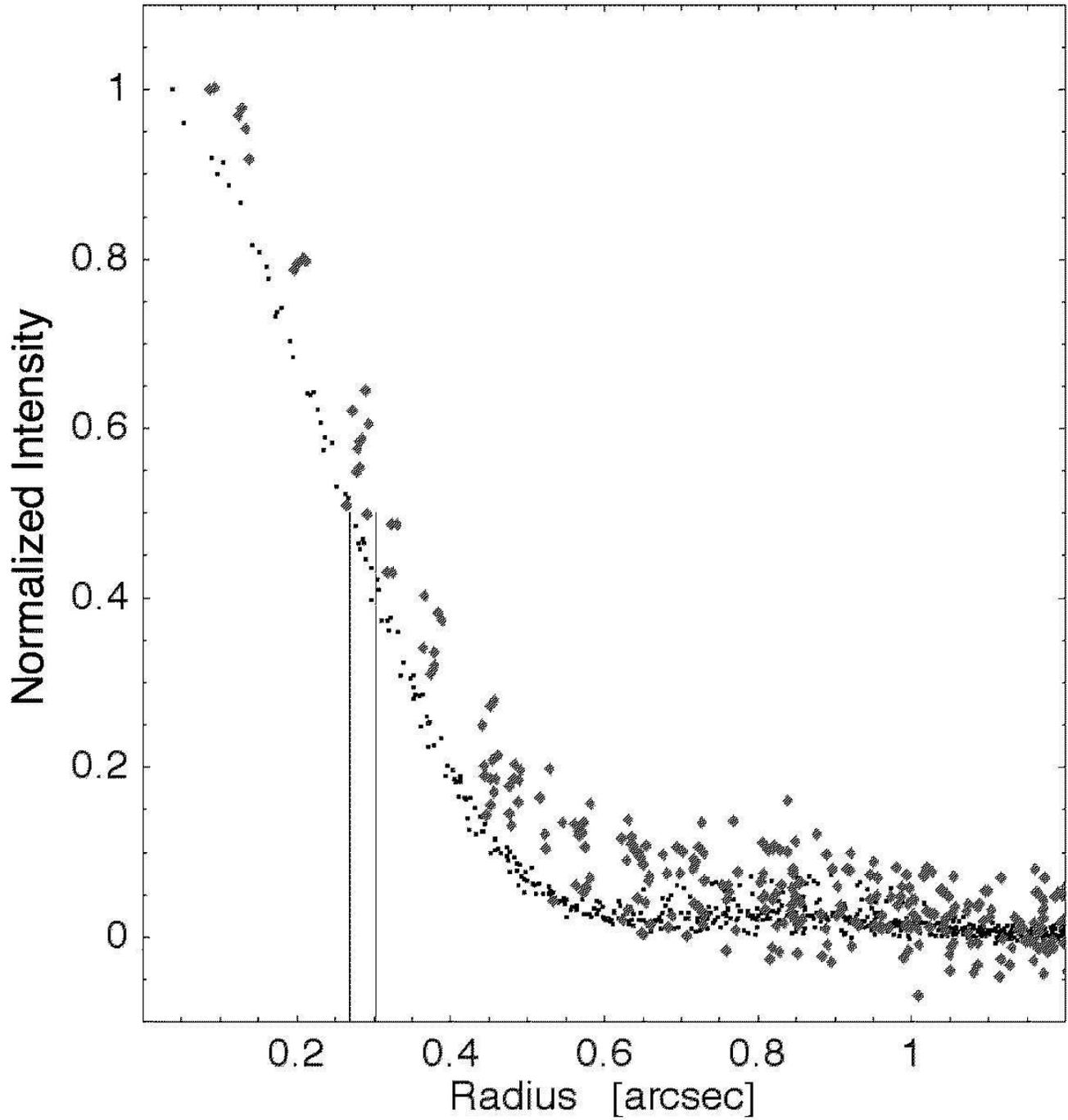}
\caption{Profiles of azimuthally averaged normalized intensity for $\zeta$ Lep (diamonds) and reference PSF star (dots); vertical lines indicate the FWHM values of profile fits to  the PSF star and  $\zeta$ Lep.  See the electronic edition of the Journal for a color version of this figure. }
\end{figure}

\end{document}